\documentstyle[preprint,aps]{revtex}
\begin{document}
\draft
\title{Classical transverse Ising spin glass with short- range interaction 
beyond the mean field approximation}
\author{}
\address{{\rm K Walasek, K Lukierska- Walasek\\ Institute of Physics, 
Pedagogical University, Plac
S{\l}owia\'nski 6, 65- 069 Zielona G\'ora, Poland\\} {\rm and} \\
{\rm L De Cesare, I Rabuffo\\ Dipartimento di Scienze Fisiche "E. R.
Caianiello" Universit\`{a} di Salerno, 84100 Salerno, and INFM Unit\`{a} di
Salerno, Italy}}
\maketitle
\begin{abstract}
The classical transverse field Ising spin- glass model with
short-range interactions is investigated beyond the mean- field
approximation for a real $d$- dimensional lattice. We use an appropriate nontrivial modification of the
Bethe- Peierls method recently formulated for the Ising spin- glass.
The zero- temperature critical value of the transverse field and the
linear susceptibility in the paramagnetic phase are obtained
analytically as functions of dimensionality $d$. The phase diagram is
also calculated numerically for different values of $d$. In the limit
$d\rightarrow\infty$, known mean- field results are consistently reproduced.
\end{abstract}
The study of glasses is today one of the most relevant and actual
problem in condensed matter physics. Originally, the basic idea was
to start from spin glass (SG) models and to extract as much as
it was possible at a mean- field- approximation (MFA) level \cite{1,2,3,4}.  
However, there are recent studies
\cite{5,6,7,8} which indicate difficulties to extend the 
MFA scenario to realistic spin glasses with short- range
interaction and decide "a priori" which properties survive and which
must be appropriately modified. Renormalization group treatments
\cite{9,10,11} for classical and quantum spin glasses and phenomenological
studies \cite{12} do not seem to suggest a clear picture. 

Quite recently, in an interesting paper \cite{13}, an approach
beyond the MFA has been achieved for an $d$- dimensional Ising SG
model with short- range interactions on a real lattice using an
extension of the Bethe- Peierls approximation (BPA) \cite{14} to the
spin glass problem via the replica trick. This approach seems to be
very promising to estabilish a direct contact with the results
obtained by different authors for the infinite- ranged version and to
controll possible deviations for short- ranged glasses from the well
acquired MFA scenario. Of course, additional applications to more
complex glassy systems and improvements are necessary for
understanding something more about the role played by the glassy
fluctuations around the MFA solution for finite dimensionalities.

In this paper we explore the glassy properties of the $d$-
dimensional classical transverse field Ising SG model \cite{15,16,17}
with short-range interactions using an appropriate nontrivial
modifications of the BPA formulated originally in Ref.
\onlinecite{13} for the Ising SG. The model here considered has
received recent attention because it is a relatively simple SG model
which reflects some properties of the quantum counterpart \cite{4}
and it is a specific example of classical two- vector anisotropic SG.
So, the classical limit of the usual more complex quantum Ising SG
model in a transverse field with realistic exchange interactions may
be useful for taking contact \cite{16,17} with low- temperature
properties of the so- called "proton glasses" \cite{15}, as the
compounds
$\text{Rb}_{1-x}\left(\text{NH}_4\right)_x\text{H}_2\text{PO}_4$, and
with the most recent experimental magnetic data for the dipolar glass
$\text{Li}\:\text{Ho}_{0.167}\text{Y}_{0.822}\text{F}_4$ \cite{18,19}.    

As concerning the quantum realistic SG model, a lot of results have
been obtained only for $d=1$ \cite{20,21} and for infinite- ranged
interactions $\left( d=\infty\right)$ \cite{4,19}.  
For the classical counterpart, only results with infinite- ranged
interaction have been derived \cite{16,17}, in particular at $T=0$,
and even in the "simple" MFA limit the full phase diagram has not yet
been calculated. In any case, there are rather little studies about
short- ranged glassy models. So, it appears quite relevant that the
BPA allows us to describe some nontrivial glassy properties of the
model (in the paramagnetic phase) for $d\geq 1$. In particular,
explicit analytical results are obtained at $T=0$ and the phase
diagram in the (temperature, transverse field)- plane is derived
numerically for an arbitrary dimensionality.

The classical transverse field SG model here considered is described
by the Hamiltonian \cite{15,16,17}:
\begin{equation} \label{1}
H=-\frac{1}{2}\sum_{\langle
i,j\rangle}J_{ij}S_i^zS_j^z-\Gamma\sum_{i=1}^NS_i^x -h\sum_{i=1}^NS_i^z
\end{equation} with
$\left(S_i^z\right)^2+\left(S_i^x\right)^2=1$.Here $\Gamma$ and $h$
are transverse and longitudinal fields, respectively, the couplings $J_{ij}$'s 
are independent random variables assuming values $\pm J$ with the 
equal probability. 
In (\ref{1}) $\sum_{\langle i,j\rangle}\cdots$ denotes
a sum over nearest- neighbours pairs of $N$ sites on a hypercubic
$d$- dimensional lattice. Using the replica trick, all is reduced to
determine the "quenched average":
\begin{equation} \label{3}
Z_n=\left[\text{Tr}\exp\left(-\beta\sum_{\alpha=1}^nH_\alpha\right)\right]_
{\text{av}}\; ,
\end{equation} where $H_\alpha$ is the $\alpha$-th replica of the
Hamiltonian (\ref{1}) and $\beta=1/T$ with the Boltzmann constant
$k_{\text{B}}\equiv 1$. Working directly on the real lattice, the
basic idea of the BPA for spin glasses \cite{13} is to take into
account the correct interactions inside replicated clusters (cl),
constitued by a central spin ${\bf S}_0$ and its $2d$ nearest-
neighbours $\{ {\bf S_i};\; i=1,\cdots ,2d\}$, and to describe the
interactions of the cluster borders with the remnant (rm) of the
system by means of effective couplings among replicas to be
determined self - consistently. With this in mind, Eq.(\ref{3}) for the 
Bethe- Peierls ansatz can be formally revritten as \cite{13}:
\begin{eqnarray} \label{4}
Z_n&=&\text{Tr}_{\{{\bf
S}_{\text{cl}}\}}\left[\exp\left(-\beta\sum_{\alpha=1}^nH_{\alpha}
^{\text{(cl)}}\right)\text{Tr}_{\{ {\bf S}_{\text{rm}}\}} 
\exp\left(-\beta\sum_{\alpha  
=1}^nH_\alpha^{\text{(rm)}}\right)\right]_{\text{av}} \\ \nonumber
&\equiv& K\left(T,\Gamma,h\right)\:\text{Tr}_{{\{\bf S}_{\text{cl}}\}}
\left[\exp \left(-\beta {\cal H}_n\right) \right]_{\text{av}}\;
\end{eqnarray} 
where \begin{equation} \label{4a} H_{\alpha}^{\text{cl}}=-\Gamma\sum_{k=0}^{2d}
S_{k\alpha}^x-\sum_{i=1}^{2d}J_{0i}S_{0\alpha}^zS_{i\alpha}^z\end{equation}
and $H_\alpha^{(\text{rm})}$ denotes replicated Hamiltonians of the cluster
and remnant of the system interacting with cluster borders,
respectively, and $K\left(T,\Gamma,h\right)$ is a multiplicative constant 
independent on lateral spins,
\begin{equation} \label{5}
\text{Tr}_{\{{\bf S}_{\text{cl}}\}}\cdots=\frac{1}{\pi^{n\left( 2d+1\right)}}
\int_{-1}^1\prod_{k=0}^{2d}\prod_{\alpha=1}^n\frac{dS_{k\alpha}^z}{\sqrt{1-
\left(S_{k\alpha}^z\right)^2}}\cdots
\end{equation} and
\begin{eqnarray} \label{6}
{\cal H}_n&=&-\sum_{\alpha=1}^n\sum_{i=1}^{2d}J_{0i}S_{0\alpha}^zS_{i\alpha}^z
-\frac{\beta J^2}{2}\sum_{\alpha,\alpha^{\prime}=1}^n\sum_{i=1}^{2d}\lambda_{
\alpha\alpha^{\prime}}S_{i\alpha}^zS_{i\alpha^{\prime}}^z \\ \nonumber
 &-&\frac{1}{\beta}\sum_{\alpha=1}^n\sum_{k=0}^{2d}\ln\cosh\left[\beta\Gamma
\sqrt{1-\left(S_{k\alpha}^z\right)^2}\right]-h\sum_{\alpha=1}^n\sum_{i=1}^{2d}
S_{k\alpha}^z\;
\end{eqnarray}with $\lambda_{\alpha\alpha^{\prime}}=\mu_{\alpha\alpha^{\prime}}
$ for $\alpha\neq\alpha^{\prime}$ and $\lambda_{\alpha\alpha}=\mu$ which are 
parameters to be determined via appropriate self- consistent equations.  
Here we have used the relation $S_{k\alpha}^x=\pm\sqrt{1-
\left(S_{k\alpha}^z\right)^2}$ $(k=0,1,\cdots.2d)$. Of course, if a transition
from a paramagnetic phase to a SG one is assumed to exist, one expects
$\mu_{\alpha\alpha^{\prime}}=0$ in the paramagnetic phase.

At this stage, the self- consistent equations which determine the
effective couplings $\mu_{\alpha,\alpha^{\prime}}$ and $\mu$ as $n\rightarrow
0$, are
\begin{equation} \label{8}
\langle S_{i\alpha}^zS_{i\alpha^{\prime}}^z\rangle =\langle S_{0\alpha}^z
S_{0\alpha^{\prime}}^z\rangle\;\; \text{with}\;\; i=1,\cdots,2d\;,
\end{equation} where 
\begin{equation}\label{10}
\langle\cdots\rangle=\frac{\text{Tr}\left[\exp\left(-\beta{\cal H}_n\right)
\cdots\right]_{\text{av}}}{\text{Tr}\left[\exp\left(-\beta{\cal H}_n\right)
\right]_{\text{av}}}\;. \end{equation} It is easy to check that, for
$h=0$, due to the inversion symmetry $S_{i\alpha}^z\rightarrow
-S_{i\alpha}^z$ and symmetry of the probability distribution for $J_{ij}$ Eq. 
(\ref{8}) with $\alpha=\alpha^{\prime}$ can be reduced 
to the following one:
\begin{equation}\label{11}
\chi_i=\chi_0\;\; \text{for}\;\; i=1,\cdots,2d \;,\end{equation}
where \begin{equation}\label{12}
\chi_k=\frac{\partial \langle S_{k\alpha}\rangle}{\partial
h}|_{h=0}\;\; \text{with}\;\; k=0,1,\cdots,2d \end{equation} denotes the
local susceptibility. 

We are now in the position to obtain the explicit equations for  
$\mu$ which will be used for obtaining also the
phase diagram of the model. Since it is expected that $\mu_{\alpha,
\alpha^{\prime}}\rightarrow 0$ approaching the spin- glass transition
from below, for $n\rightarrow 0$, one obtains at $h=0$
\begin{eqnarray} \label{13}
\langle S_{k\alpha}^zS_{k\alpha^{\prime}}^z\rangle&=&
\beta J^2\mu_{\alpha\alpha^{\prime}}\sum_{i=1}^{2d}\left[\langle S_k^zS_i^z\rangle_0
\right]_{\text{av}}+{\cal O}\left(\mu_{\alpha,\alpha^{\prime}}^2\right)\\ \nonumber
& & \left(k=0,1,\cdots,2d\right) \;,\end{eqnarray}
where \begin{equation}\label{14}
\langle\cdots\rangle_0=\frac{1}{\pi^{2d+1}Z_0}\int_{-1}^{1}\prod_{k=0}^{2d}
\frac{dS_k^z}{\sqrt{1-\left(S_k^z\right)^2}}\exp\left(-\beta{\cal H}_0\right)
\cdots \end{equation} with \begin{equation}\label{15} {\cal
H}_0=-\sum_{i=1}^{2d}J_{0i}S_0^zS_i^z-\frac{\beta J^2}{2}\mu\sum_{i=1}^{2d}
\left(S_i^z\right)^2-\frac{1}{\beta}\sum_{k=0}^{2d}\ln\cosh\left[\beta\Gamma
\sqrt{1-\left(S_k^z\right)^2}\right]\;.
\end{equation} In Eq. (\ref{14}) $Z_0$ denotes the normalization factor. The 
term $-h\sum_{k=0}^{2d}S_k^z$ must be added to the right hand side of
(\ref{15}) when it is necessary. So, at
$h=0$ due to the translational symmetry for the sample averaged
system and assuming $\mu_{\alpha\alpha^{\prime}}=0$ at and above the
glassy transition line (to be determined), the self- consistent
equation (\ref{8}) for ${\alpha\neq \alpha^{\prime}}$ and $\alpha=\alpha^
{\prime}$ reduces ,respectively, to:
\begin{eqnarray}\label{17}
\left[\langle\left(S_i^z\right)^2\rangle_0^2\right]_{\text{av}}+&\left(2d-1
\right)\left[\langle S_i^zS_j^z\rangle_0^2\right]_{\text{av}}=2d\left[\langle
S_0^zS_j^z\rangle_0^2\right]_{\text{av}}\\ \nonumber
&(i\neq j = 1,\cdots,2d)
\end{eqnarray} and
\begin{equation}\label{18}
\left[\langle\left(S_i^z\right)^2\rangle_0\right]_{\text{av}}=\left[\langle
\left(S_0^z\right)^2\rangle_0\right]_{\text{av}}\;\;\;(i=1,\cdots,2d)\;,
\end{equation} where $i\neq j$ denote arbitrary lateral sites of
the cluster with the central spin ${\bf S}_0$.

By solving Eqs (\ref{17})- (\ref{18}), it is possible to obtain the
phase diagram of our model in the $\left(T,\Gamma\right)$ plane.
Explicit results can be derived analytically only at
$T=0$.  As
$T\rightarrow 0$, introducing $\overline{\mu}=\beta \mu$ which is
finite, with the help of Eq. (\ref{14}) 
choosing in Eq. (\ref{18}) $i=1$ one obtains:
\begin{equation}\label{21}
\overline{\mu}=\frac{1}{J^2}\frac{\Gamma-\sqrt{\Gamma^2-4\left(2d-1\right)J^2}
}{2} \end{equation} and hence taking into account Eq. (\ref{11}) one finds, in 
the paramagnetic phase at $T=0$,  linear susceptibility:
\begin{equation}\label{22}
\chi=2\frac{\Gamma+\sqrt{\Gamma^2-4\left(2d-1\right)J^2}}{\left(\Gamma+
\sqrt{\Gamma^2-4\left(2d-1\right)J^2}\right)^2-4J^2}\;\;.\end{equation}
Now, we calculate the critical value $\Gamma_c$ 
 at $T=0$ of the transverse field using Eqs (\ref{17}) and (\ref{21}). 
For $i=1$ and $j\neq 1$ , with some
algebra we rewrite Eq.(\ref{17}), as $T\rightarrow 0$, in the
following form:
 \begin{equation}\label{25}
\left(\Gamma_c-\overline{\mu}_c J^2\right)^4-2dJ^2\left(\Gamma_c-\overline{\mu}
_cJ^2\right)^2+\left(2d-1\right)J^4=0 \end{equation} with $\overline{\mu}_c
J^2=\frac{1}{2}\left[\Gamma_c-\sqrt{\Gamma_c^2-4\left(2d-1\right)J^2}\right]$.
From this equation one easily obtains:
\begin{equation}\label{26}
\Gamma_c=2\left(2d-1\right)^{1/2}J\;.\end{equation} As we see, $\chi$ is
positive and has physical meaning only for $\Gamma\geq \Gamma_c$.
This suggests that the expression (\ref{22}) for $\chi$ is related
only to the paramagnetic phase. 

For $\left(\Gamma-\Gamma_c\right)/\Gamma_c\ll 1$ Eq.(\ref{22}) yields:
\begin{equation}\label{28}
\chi\approx\left\{\begin{array}{ccl}
\frac{\left(2d-1\right)^{1/2}}{2J\left(d-1\right)}\left[1-\frac{d\sqrt{2}}
{d-1}\left(\frac{\Gamma-\Gamma_c}{\Gamma_c}\right)^{1/2}+{\cal O}\left(\frac{
\Gamma-\Gamma_c}{\Gamma_c}\right)\right] & \text{for} &d\neq 1\\
\frac{1}{2J\sqrt{2}}\left(\frac{\Gamma-\Gamma_c}{\Gamma_c}\right)^{-1/2}
\left[1+{\cal
O}\left(\frac{\Gamma-\Gamma_c}{\Gamma_c}\right)^{1/2}\right] & \text{for} &
d=1 \end{array}\right. \;. \end{equation}. As an "a posteriori"
justification of the correctness of the glassy BPA (\ref{4}), it is
easy to check analytically that, using the rescaling $J\rightarrow
J/\sqrt{2d}$ one finds $\overline{\mu}=2d\chi$ and we get at $T=0$
for $d\rightarrow\infty$; $ \chi=\frac{1}{2J^2}\left[\Gamma-
\left(\Gamma^2-\Gamma_c^2\right)^{1/2}\right]$ with $\Gamma_c=
2J+{\cal O}\left(d^{-1/2}\right)$. These results reproduce
exactly those obtained at $T=0$ for the same SG model but with
infinite- ranged interactions \cite{16}. This partial result supports
the validity of the BPA for SG's.

The situation for $d=1$ with a divergence of the linear
susceptibility at $T=0$ as $\Gamma\rightarrow \Gamma_c^+$ can be
simply explained. With the dichotomic probability distribution of one
-dimensional nearest- neighbours couplings $J\left(i,i+1\right)\equiv
J_i$, $\left(J_i=\pm J\;\; \rm{with}\;\; J>0\right)$ after the
gauge transformation of spin variables \begin{equation} \label{28a}
S_i^z\rightarrow \text{sign}\left(J_1\right)\cdots \text{sign}\left(J_{i-1}\right)
S_i^z\end{equation}the sytem can be reduced to the uniform ferromagnet in
an external  transverse field $\Gamma $. Therefore it is naturally to
expect that at $\Gamma=\Gamma_c$ the ferromagnetic phase transition
with a divergent linear susceptibility occurs. Indeed a more detailed
analysis of the one dimensional case shows that at $T=0$ the linear
susceptibility $\chi$ can be calculated exactly for the paramagnetic
phase. The divergence of $\chi$ is the same as that obtained within
the BPA for the one- dimensional system. 

For arbitrary $T$ and $\Gamma$, from Eqs
(\ref{17}) and (\ref{18}) one can calculate numerically equilibrium 
properties ofour model in the paramagnetic phase. 
In particular in Fig.\ref{fig1}, the phase diagram in the
$(\Gamma,T)$- plane for different $d$ is shown. We have conveniently
scaled variables $T$ and $\Gamma$ for reproducing results at very
high dimensionality. In Fig. \ref{fig2} a variation of rescaled
critical temperature with a dimension at $\Gamma=0$ is plotted.

In conclusion, we have studied some relevant aspects
of the classical transverse field short- ranged Ising SG in the
paramagnetic phase for arbitrary dimensionality $d$. We expect that
our results may be also useful for explaining some properties of the
quantum counterpart of the model here considered. However, some
questions remain to be explained. For example, on the basis of the general 
self- consistent equation (\ref{8}) it is 
interesting to find solutions with $\mu_{\alpha\alpha
^{\prime}}\neq 0$ in order to see if the BPA is able to describe
correctly our model in the SG phase at arbitrary dimensionalities.
Within present calculations working for paramagnetic phase this is
practically impossible, since the complicated integral (\ref{14}) has
been reduced at $T=0$ to the asymptotic form being Gaussian like one.
Such an asymptoptic form is insufficient when parameters $\mu_{\alpha,
\alpha^{\prime}}$ are included even in the replica symmetric form. 
Therefore further works will be necessary to elucidate these problems.

Authors thank to Professor M. Fusco Girard for numerical
calculations. Discussions with Professor Th. M. Nieuwenhuizen, Drs R. Monasson 
and R. Zecchina are appreciated.  Two of us (K. W. and K. L- W.) would like 
to express our thanks for the Department of Theoretical Physics of Salerno
University shown to us during the preparation of this paper.
Additional support from Polish Committee for Scientific Research (K.
B. N.), Grant No 2 P03B 034 11 is gratefully acknowledged. 

\begin{figure}
\caption{Phase diagram of the classical transverse
Ising spin- glass with short- range interaction within the Bethe-
Peierls approximation for spatial dimensions  $d=2,3,4,5$ and 
 $6$. The temperature  $T$ and transverse field  
$\Gamma $ are rescaled by the
factor $\left(2d\right)^{-1/2}$ . The larger the dimension, the higher the
corresponding line. Here $J\equiv 1$.}  \label{fig1} 
 \end{figure}
\begin{figure} 
\caption{Rescaling critical temperature $T_c\left(2d\right)^{-1/2}$
  for  $\Gamma =0$ versus the dimension $d$. Here $J\equiv 1 $.}\label{fig2}
\end{figure}
\end{document}